\documentclass[aps,prab,twocolumn,superscriptaddress,groupedaddress]{revtex4}

\usepackage{hyperref}
\hypersetup{
    colorlinks=true,
    linkcolor=green,
    filecolor=blue,
    urlcolor=blue,
}
\usepackage[normalem]{ulem}
\usepackage{mathtools}

\usepackage{lineno}
\usepackage{amsmath}
\usepackage{graphicx}
\usepackage{caption}
\usepackage{subcaption}
\usepackage{color}
\usepackage{natbib}
\usepackage{comment}
\usepackage{dcolumn}
\usepackage{bm}
\usepackage{bbm}

\begin{document}

\title{Uncertainty Quantification
for Virtual Diagnostic of Particle Accelerators}
\author{Owen Convery}
\affiliation{SLAC National Accelerator Laboratory, Menlo Park, CA 94025, USA}
\author{Lewis Smith}
\affiliation{Department of Computer Science, University of Oxford, UK}
\author{Yarin Gal}
\affiliation{Department of Computer Science, University of Oxford, UK}
\author{Adi Hanuka}\thanks{Corresponding author: adiha@slac.stanford.edu}

\affiliation{SLAC National Accelerator Laboratory, Menlo Park, CA 94025, USA}

\date{\today}
\begin{abstract}

Virtual Diagnostic (VD) is a computational tool based on deep learning that can be used to predict a diagnostic output.
VDs are especially useful in systems where measuring the output is invasive, limited, costly or runs the risk of altering the output.
Given a prediction, it is necessary to relay how reliable that prediction is, i.e. quantify the uncertainty of the prediction. In this paper, we use ensemble methods and quantile regression neural networks to explore different ways of creating and analyzing prediction's uncertainty on experimental data from the Linac Coherent Light Source at SLAC National Lab. We aim to accurately and confidently predict the current profile or longitudinal phase space images of the electron beam.
The ability to make informed decisions under uncertainty is crucial for reliable  deployment of deep learning tools on safety-critical systems as particle accelerators.
\end{abstract}

\maketitle
\section{Introduction}

Particle accelerators serve a wide variety of applications ranging from  chemistry, physics to biology experiments. Those experiments require increased accuracy of diagnostics tools to measure the beam properties during its acceleration, transport and delivery to users. Diagnostics must keep pace with the advance of extreme beam conditions and the increased experiments' complexity, which presents challenges to the current state-of-the-art \cite{ XTCAV3,XTCAV4}.
Machine learning (ML) applications for accelerator diagnostics have been recently shown to improve beam stability \cite{ALS:LEEMAN}, identify glitchy accelerator components \cite{Tennant_glitch}, perform optics corrections and detect faulty beam position monitors \cite{fol_bpm, ARPAIA:LHC}, and control in real-time by combining adaptive feedback and ML \cite{ES:NN}.

Given readily available input data, virtual diagnostic (VD) tools provide a shot-to-shot non-invasive measurement of the beam in cases where the diagnostic has limited resolution or limited availability \cite{Emma2018, hanuka2020accurate, FELVD}. VDs have the potential to be useful in experiment's design, setup and optimization while saving valuable operation time. They could also aid in  interpreting experimental results, especially in cases in which current diagnostics cannot provide necessary information.

Current VD provides predictive models based on training a neural network mapping between non-invasive diagnostic input to invasive output measurements \citep{Emma2018, hanuka2020accurate,  Emma2019IBIC}. This type of mapping is known as supervised regression. Previous work has demonstrated VD to predict the electron beam current profile and Longitudinal Phase Space (LPS) distribution \cite{DESY:lps} along the accelerator using either scalar controls \cite{Emma2018} or spectral information \cite{hanuka2020accurate} as the non-invasive input to the VD.
However, while current VD methods provide mean prediction only, an essential component in deploying the virtual diagnostic tool is to quantify the confidence in the prediction, i.e. estimate an interval presenting the uncertainty in prediction. 
A metric of uncertainty and its reliable presentation provide a way of making informed decisions, that becomes crucial in safety-critical systems such as particle accelerators.

In general, there are two classes of uncertainty: epistemic and aleatoric uncertainty \cite{kendall2017uncertainties}. Aleatoric, or statistical irreducible uncertainty, is the uncertainty in the data set that arises from experimental error or inherent measurement noise. Given the same set of inputs, we may observe slightly different results. Epistemic uncertainty, also known as systematic reducible uncertainty, is produced when the model's knowledge is limited or hindered, e.g. from missing data. We aim to capture both types of uncertainty in the VD tool and to incorporate them into the final prediction. While the neural network can only make point predictions about beam properties, we can use tools from deep learning to better understand the uncertainty of the predictions.

There are various ways to quantify uncertainty of machine learning models for supervised regression problems.
These include Bayesian model averaging approaches such as Gaussian processes \cite{rasmussen2003gaussian} and Bayesian neural networks (BNNs) \citep{mackay1992bayesian}, which learn distributions over the parameters of the network. Compared to Gaussian processes, BNNs have the advantage that they can easily scale to high dimensional inputs. However, BNNs are computationally expensive and require modifications to the training procedure. Other non-Bayesian approaches widely used are bootstrapping, ensembling \cite{bickel1981some, clarke2003comparing} and quantile regression \cite{qr,qr:ALD}. Those non-Bayesian approaches are simple to implement, could be easily parallelized to scale with large amount of data, and yields high quality predictive uncertainty estimates \cite{lakshminarayanan2017simple,ensemble_review}. Ensembles are closely related to Bayesian model averaging, and ensembles of neural networks can be seen as an ad hoc approximation of Bayesian neural networks or Gaussian processes for specific architectures and training conditions \cite{gal2016uncertainty, smith2018understanding, he2020bayesian}.
When each ensemble component is trained with random initialization \emph{and} a random data subset, the procedure is known as bootstrap aggregation, or bagging. Bootstrapping is well studied in the Frequentist statistics literature and is closely related to Bayesian statistics with certain priors \citep{efron1982jackknife, efron1994introduction, bickel1981some}. 

In this paper, we apply deep learning tools to provide confidence intervals for virtual diagnostic predictions.
We compare the predicted uncertainty learned by multiple ensemble methods as well as quantile regression using experimental data from the Linac Coherent Light Source (LCLS) at SLAC National Lab \cite{emma:lcls}.
We evaluate the robustness of the ensemble methods to provide accurate mean predictions as well as reliable uncertainty estimation.

The paper is organized as follows: in \autoref{sec:methods} we first present the VD architecture. Then we discuss metrics to evaluate the reliability of the prediction to be used online on the particle accelerator. In \autoref{sec:results} we demonstrate the methods and compare them on two experimental datasets from LCLS; 1D current profiles and 2D longitudinal phase space images. The 1D current profile is a projection of the LPS image. Finally, in \autoref{sec:discuss} we discuss re-weighing the data sets, and other neural network architectures to improve the results.

\section{Methods\label{sec:methods}}

In this section we first present the Virtual Diagnostic (VD) neural network architecture and the data sets. We then present various ensemble and quantile methods to quantify the uncertainty. Last, we present the metrics used to evaluate the accuracy of the VD's mean prediction and uncertainty.

\subsection{VD Architecture and Data Set \label{sec:NN}}

\begin{figure}[!h]
    \centering
    \includegraphics[scale=0.5]{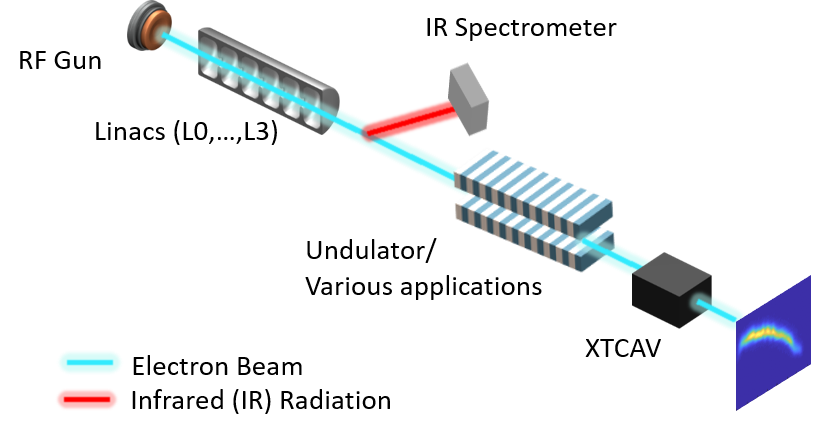}
    \caption{Schematic of a generic, linac-driven light source.} Reproduced from \cite{hanuka2020accurate}.
    \label{fig:accelerator}
\end{figure}

High brightness beam linacs typically operate in single-pass, multi-stage configurations where a high-density electron beam formed in the RF gun is accelerated and manipulated prior to delivery to users in an experimental station. An example of such a facility is the LCLS XFEL at SLAC where the electron beam traverses through an undulator, and emits coherent X-ray pulses. Typically, longitudinal phase space (LPS) is destructively measured by an X-band transverse deflecting cavity (XTCAV) \cite{XTCAV, XTCAV3} - as shown in \autoref{fig:accelerator}.
  
\begin{figure}[!htpb]
    \centering
\begin{subfigure}{.23\textwidth}
    \centering  \includegraphics[width=3.85cm]{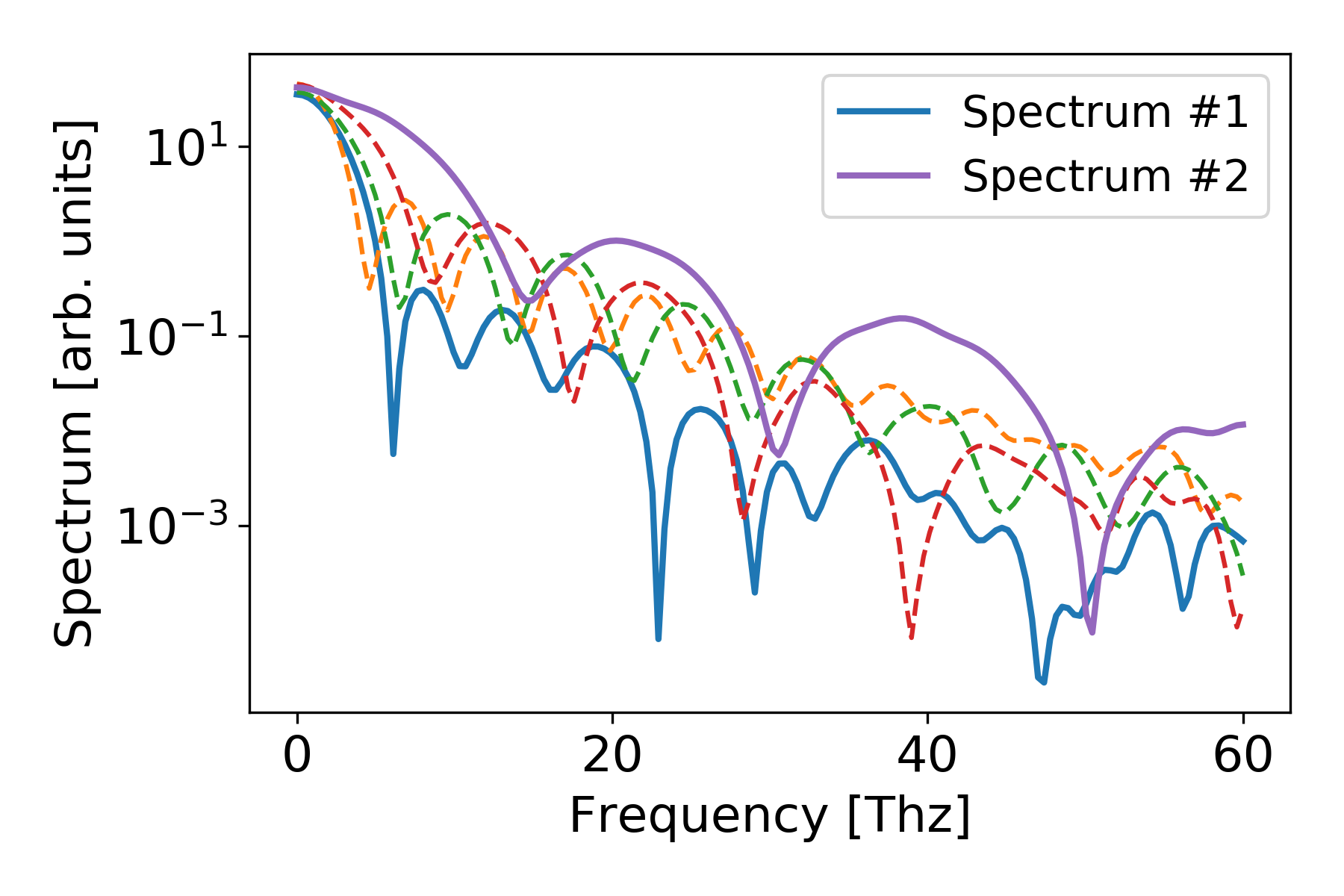} 
\caption{Spectra}   
\end{subfigure}
    \begin{subfigure}{.23\textwidth}
    \centering  \includegraphics[width=3.85cm]{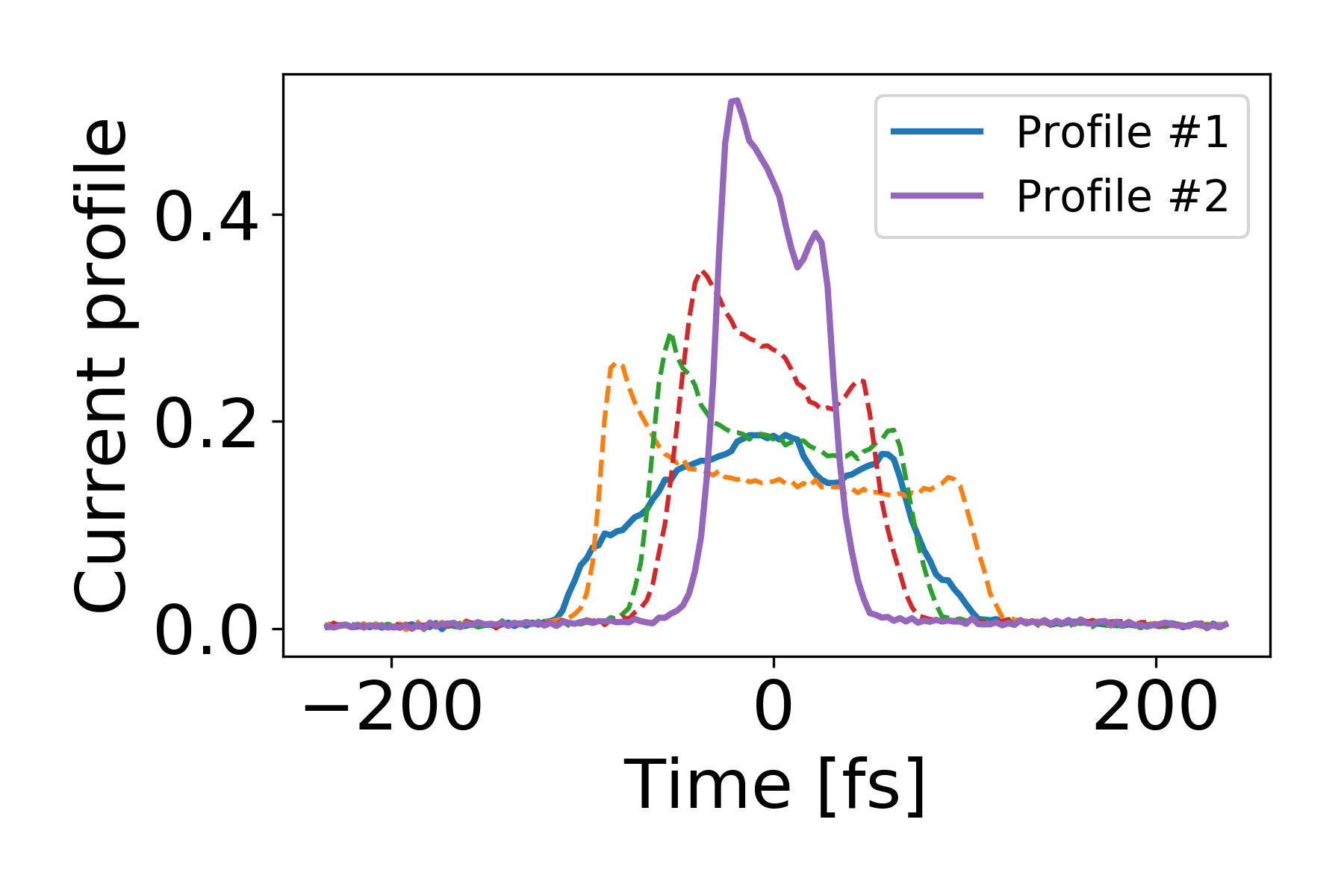} 
\caption{Current Profiles} 
\label{fig:profile_ex}
\end{subfigure}
    \begin{subfigure}{.23\textwidth}
    \centering  \includegraphics[width=3.85cm]{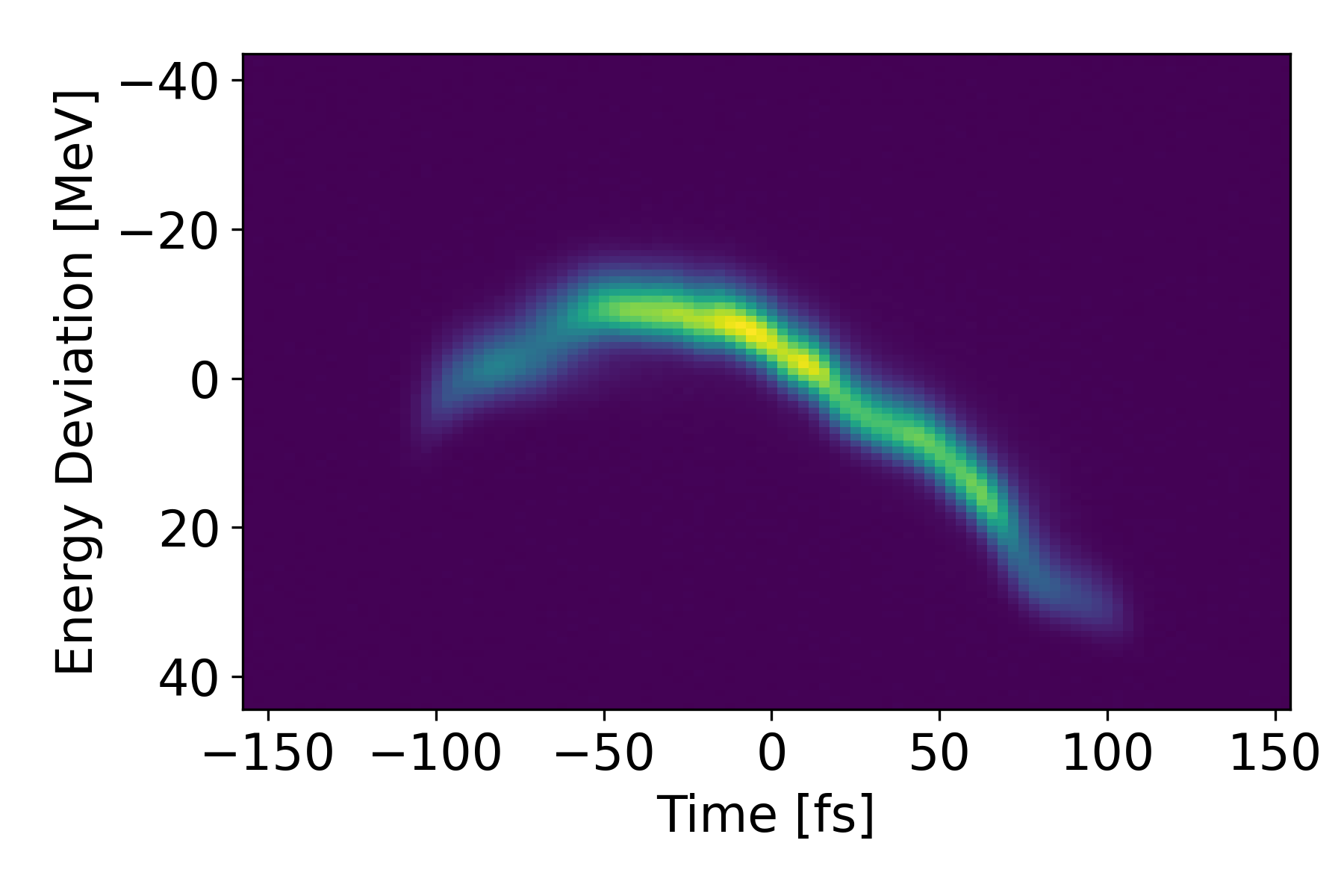} 
\caption{ LPS for spectrum $\#$1} 
\label{fig:lps_ex01}
\end{subfigure}
    \begin{subfigure}{.23\textwidth}
    \centering  \includegraphics[width=3.85cm]{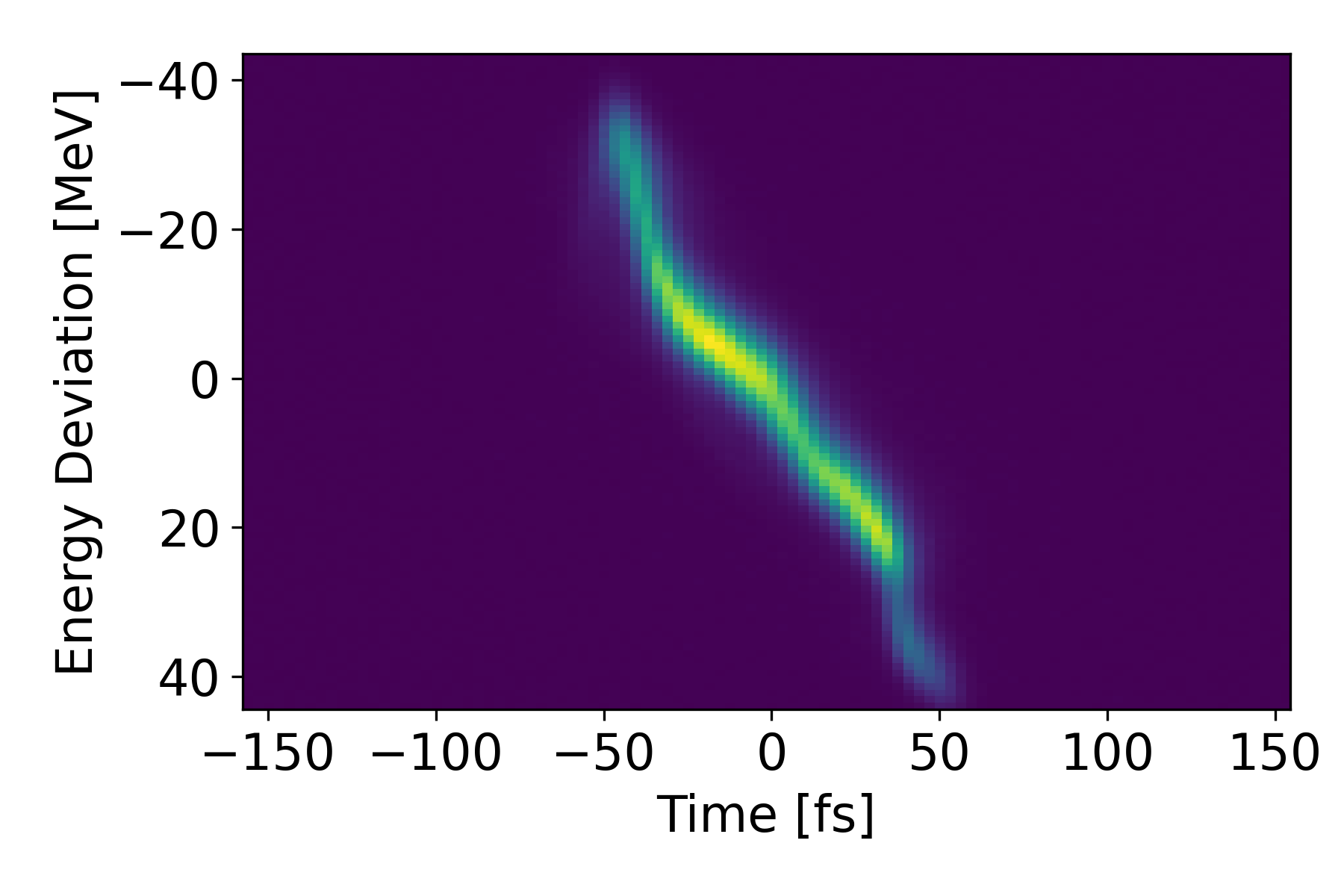} 
\caption{ LPS for spectrum $\#$2} 
\label{fig:lps_ex02}
\end{subfigure}
    \caption{Example of VD input of spectrum in (a) and corresponding outputs; either 1D current profiles in (b), or images of Longitudinal Phase Space (LPS) --- examples of which are shown in (c) and (d) for two different spectra input.}%
    \label{fig:VD_input_output}%
\end{figure}

In this paper we used two experimental data sets: 2D LPS images or 1D current profiles, measured at the XTCAV, as the outputs, and the corresponding spectral information, as can be collected by IR spectrometer, as an input.
The 1D current profile is a projection of an LPS image.
The outputs have been centered and cropped about their region of interest as in Ref \cite{Emma2018}.
 Examples of the inputs and outputs are shown in \autoref{fig:VD_input_output}. 
Both data sets contain 4,046 shots and were randomly shuffled and split 80$\%$ for training and validation, and 20$\%$ for testing. We then normalized all output profiles and images. The outputs shown in the paper are all normalized.

The neural network (NN) architecture we used is a fully connected feed-forward NN composed of three hidden layers (200, 100, 50) with rectified linear unit activation function.
Other network architectures are discussed in \autoref{sec:discuss}. For training we used a batch size of 32, 500 epochs and the Adam optimizer with a fixed learning rate of 0.001 \cite{hanuka2020accurate}. 
Training the NNs with a Gaussian likelihood, i.e. to minimize the standard Mean Squared Error (MSE) loss function on a training set, yields symmetric uncertainty intervals. 
To introduce asymmetric uncertainty intervals relative to the mean value, we trained on quantiles of data as described in \autoref{sec:uqmethods}. 
The open source Keras and TensorFlow libraries were used to build and train the NN module \cite{keras,TF}. 

\subsection{Uncertainty Quantification Methods\label{sec:uqmethods}}

Various ways exist to estimate an interval presenting the uncertainty in the neural network prediction. In what follows, we describe a few ensemble methods, wherein a collection of various neural networks is combined together. In this case, the variance of the ensemble is an estimate of the uncertainty.
We then describe quantile regression to predict multiple quantiles of the data. We also applied the popular MC dropout technique \cite{gal2016_dropout}. However, we observed a degraded mean prediction and inaccurate uncertainty bounds compared to ensembling across the entire dataset. Further improvement could combine ensembles and MC dropout but this is beyond the scope of this paper.

\subsubsection{Ensemble methods\label{sec:ens}}

Deep ensembles are ensembles of neural networks, each initialized differently and optimized to converge independently of the others. 
Ref. \cite{lakshminarayanan2017simple} has shown empirically that explicit ensembling {can} result in improved uncertainty estimates when they used large neural networks with non-convex loss surfaces.
Deep ensembles are closely related to Bayesian neural networks and Gaussian processes. For example, Ref. \cite{he2020bayesian} has recently shown that with a small adaptation to their training procedure, deep ensemble components can be interpreted as draws from a Gaussian process posterior with a certain kernel.

We investigated three ways of generating this ensemble: using random initialization of the NN parameters, using a random subset of the data (e.g. k-fold cross validation), and bagging (e.g. using both).
The predicted current profile for a test shot $\vec{I}_{{\rm predicted}}=M^{-1}\sum_{m=1}^M \vec{i}_{{\rm predicted},m} $ is the mean prediction of $M$ neural network predictions $\vec{i}_{{\rm predicted},m} $.
The uncertainty is manifested as the standard deviation of the neural network predictions $\vec{\sigma}=\sqrt{M^{-1}\sum_{m=1}^M (\vec{i}_{{\rm predicted},m}-\vec{I}_{\rm predicted})^2}$.

{\textbf{Random parameters initialization.}}
Random initialization of the neural network parameters (e.g. weights) is a promising approach for improving prediction accuracy \cite{lee2015m} and uncertainty \cite{lakshminarayanan2017simple}. 
It has been recently shown that random initializations explore entirely diverse solutions in function space \cite{fort2020deep} rather than collapse to the same solution \cite{gal2016uncertainty}. 
Popular initializations used in many applications are the He and Glorot uniform distributions \cite{HE:ICCV,glorot}. Those initializations often lead to quick and reliable convergence during training. However, to create a more diverse ensemble of neural networks, a random normal initialization has the potential to reduce the amount of near identical solutions. Here, we initialized all NN parameters as independent and identically distributed random variables from $\mathcal{N}(0,0.05)$ when training on the same predetermined training/validation data split. Glorot Uniform is used for initialization in all other parts of this paper.

{\textbf{K-Fold cross validation}}
Rather than using the same training/validation split for each NN in the ensemble, we randomly select a subset of the data. We fixed the seed for a Glorot uniform distribution for all ensemble components. We split the training data set into $K$ partitions for $K$ neural networks. Using a different partition for validating each model would yield a global model that can generalize better since it has been trained/validated using many subsets of the same data. 
This is a more responsible way to validate models for ensembles \cite{raschka2018model}.

\begin{figure}[h!]
    \begin{subfigure}{0.48\textwidth}
    \includegraphics[scale=0.4]{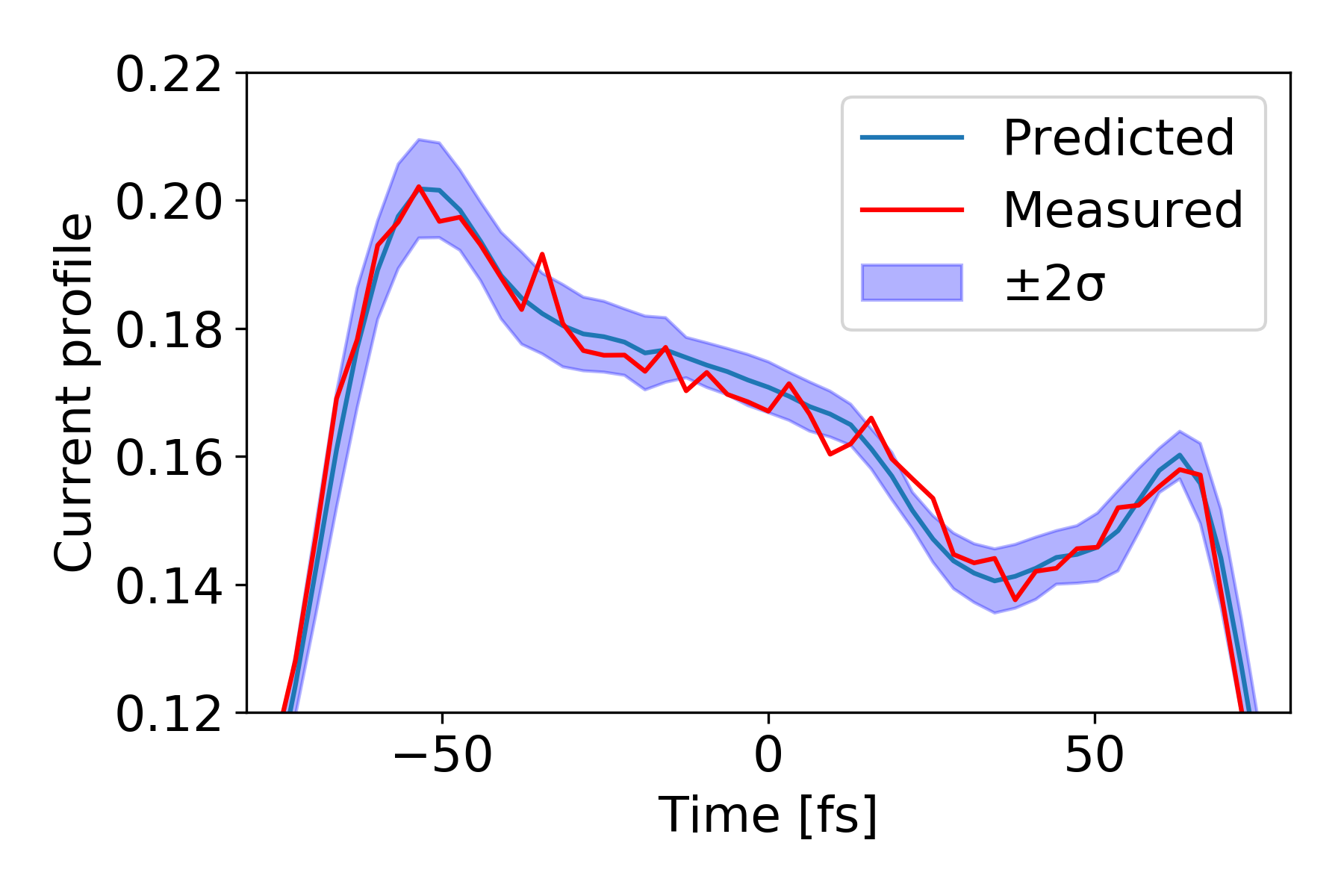}
    \caption{Bagging ensemble}
    \label{fig:bagging}
     \end{subfigure}
\begin{subfigure}{0.48\textwidth}
    \centering
    \includegraphics[scale=0.4]{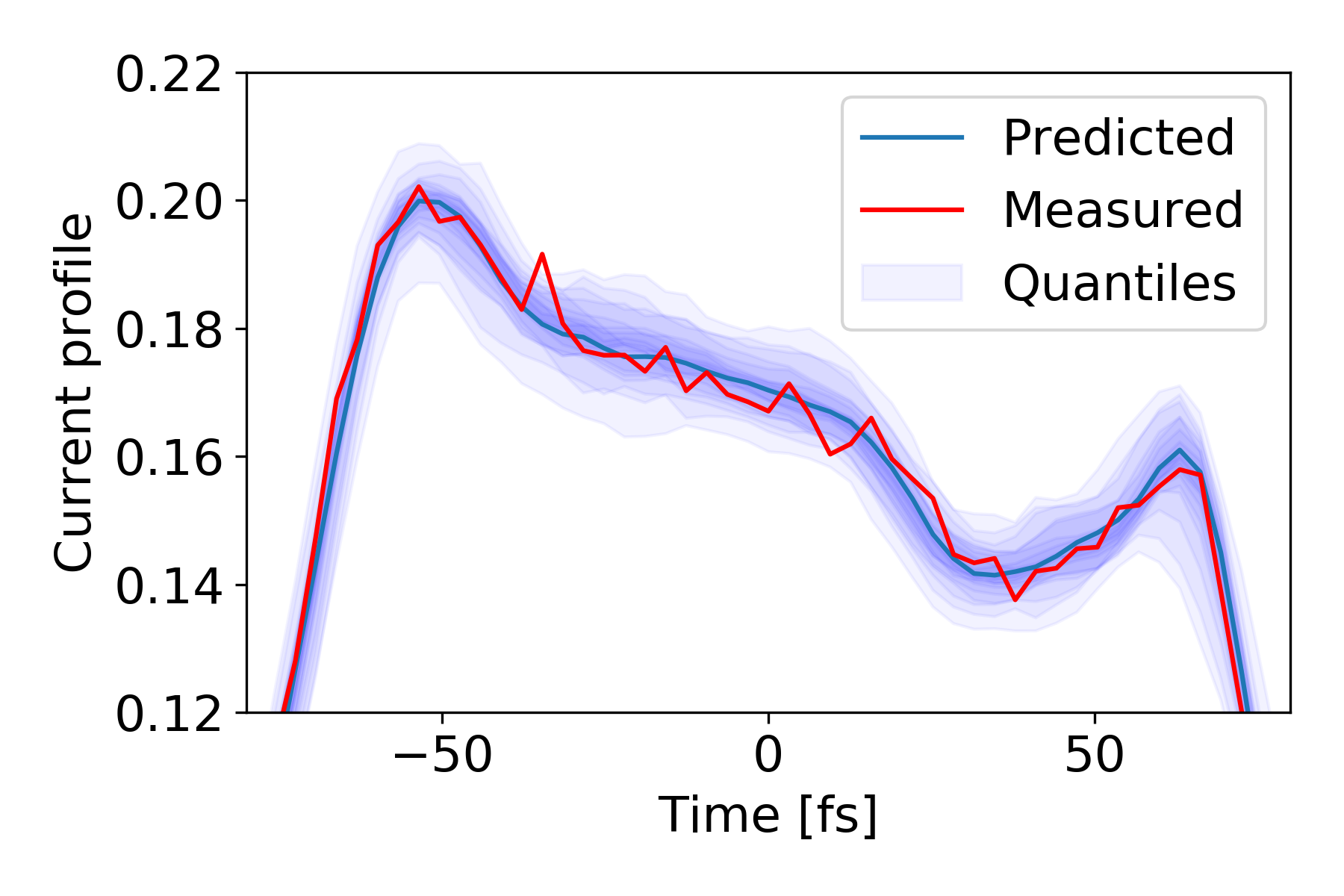}
    \caption{Quantile regression}
    \label{fig:quantile_ex}
    \end{subfigure}
    \caption{Normalized measured current profile (red) of shot $\#729$ and its prediction (blue) using (a) bagging with 16 ensembles and with $\pm2\sigma$ uncertainty interval, and (b) quantile regression with 90$\%$ uncertainty interval.}
\end{figure}

{\textbf{Bagging.}} Bootstrap AGGregatING (bagging) is an ensemble method that trains each NN on a different `bag' of data. Each bag randomly (with replacement) contains $n'$ out of $n$ possible data points where $n'/n$ is typically $\sim60\%$. All ensemble components are initialized with a different Glorot uniform seed, resulting in a NN model with both random sub-sampling of the data and random initialization. Bagging is a method commonly used to reduce variance and avoid over-fitting \cite{Opitz_1999}. \autoref{fig:bagging} shows an example of the measured current profile (normalized) $\vec{I}_{\rm measured}$, its mean prediction $\vec{I}_{\rm predict}$ and the uncertainty ($\pm2\vec{\sigma}$) of bagging with 16 ensemble components.

\subsubsection{Quantile regression\label{sec:qr}}

Neural networks can be trained to predict multiple quantiles of the data \cite{qr} by using a tilted loss function 
\begin{equation}
 { {\rm\mathcal{L}}(\zeta_t|\tau)} =
    \begin{cases}
        \tau \zeta_t & \text{if $\zeta_t \geq 0$} \\
        (\tau - 1)\zeta_t & \text{if $\zeta_t < 0$} \\
  \end{cases} 
  \label{eq:qunatile_loss}
\end{equation}
where ${{\zeta_t} }={I}_{{\rm measured},t}-{I}_{{\rm predicted},t}$ for each time index $t$. The average loss over the entire shot of length $T$ is ${T^{-1}{\sum}_{t=1}^{T}{\mathcal{L}}(\zeta_t|\tau)}$.
In practice, a separate NN is used to predict each quantile. Each is trained using the tilted loss function but with a different $\tau$ value corresponding to the desired quantile, 0.5 being the median prediction.

\autoref{fig:quantile_ex} shows 19  quantiles between 0.05 and 0.95. When plotting multiple quantile lines at once, we can gain better insight into where the ground truth may lie.
We used the median quantile $(\tau=0.50)$ for calculating the MSE of the prediction.

\subsection{Metrics for Evaluation\label{sec:metrics}}

In order to compare different approaches to quantify uncertainty, we need to have a way to evaluate the desirable characteristics of the uncertainty. Those metrics of the predictive uncertainty quality will be used in real-time during the machine operations as well to indicate the validation of the VD's prediction.  We used a custom accuracy metric to evaluate the model's performance in capturing the ground truth. At each time index ($t$) in the shot, we check whether the measured ground truth value ($I_{{\rm measured},t}$) was within an upper $I_{{\rm upper},t}$ and lower $I_{{\rm lower},t}$ prediction bounds. We then weight it by the measured value to penalize the near-zero noise while focusing on important features in the signal. 
The prediction accuracy of a shot with a length $T$ is defined as:

\begin{equation}
    {{\rm Accuracy }=  \frac{\sum_{t=1}^{T}\alpha_t\cdot  I_{{\rm measured},t}^{2}}{ \sum_{t=1}^{T} I^{2}_{{\rm measured},t}}}
        \label{eq:accuracy}
\end{equation}
 
where ${\rm \alpha}_t =\mathbbm{1}(I_{{\rm lower}{,t}} < I_{{\rm measured}{,t}} < I_{{\rm upper}{,t}})$. Unless otherwise stated, for a symmetric loss function we used bounds of $I_{{\rm upper}{,t}}, I_{{\rm lower}{,t}} = I_{{\rm predicted}{,t}}\pm2\sigma_{t}$ where $\sigma_t$ is predictive standard deviation at time $t$. For the tilted loss function, the NN predictions using $\tau=0.05$ and $\tau=0.95$ were used for $I_{\rm lower}$ and $I_{\rm upper}$ respectively. We chose to use $\pm2\sigma$ of the ensembles and $0.90$ $(0.95-0.05)$ of the quantiles for these calculations because we wanted to capture $\geq90\%$ of the uncertainties correlated with each method.

Since the measured ground truth will not be available when operating the machine in real-time, in order to to discern a good prediction from a poor prediction we correlate between the mean squared error between the measured and predicted values - ${\rm MSE} = {T^{-1}\sum_{t=1}^{T} {(I_{{\rm measured},t}-I_{{\rm predicted},t})}^{2}}$ - for a vector length of $T$ and the maximum predicted uncertainty $\sigma_{\rm max}=\max\{\vec \sigma\}$. A positive correlation indicates that the model can accurately inform which predictions have low vs. high uncertainty before seeing the ground truth. If choosing to retain/reject predictions in practice, a retention curve can be used to set a proper threshold for deployment; we retain (and predict) shots with maximum standard deviation per shot smaller than a given threshold. 

\section{Results \label{sec:results}}

In this section, we evaluate the methods described in \autoref{sec:uqmethods} with the metrics discussed in \autoref{sec:metrics} on the 1D current profiles. Next, we apply those techniques to the 2D LPS image data set. Finally, we discuss problems that are unique to the images and how they can be relieved.

\subsection{1D Current Profiles}

\begin{figure}[h!]
    \centering
      \includegraphics[scale=0.4]{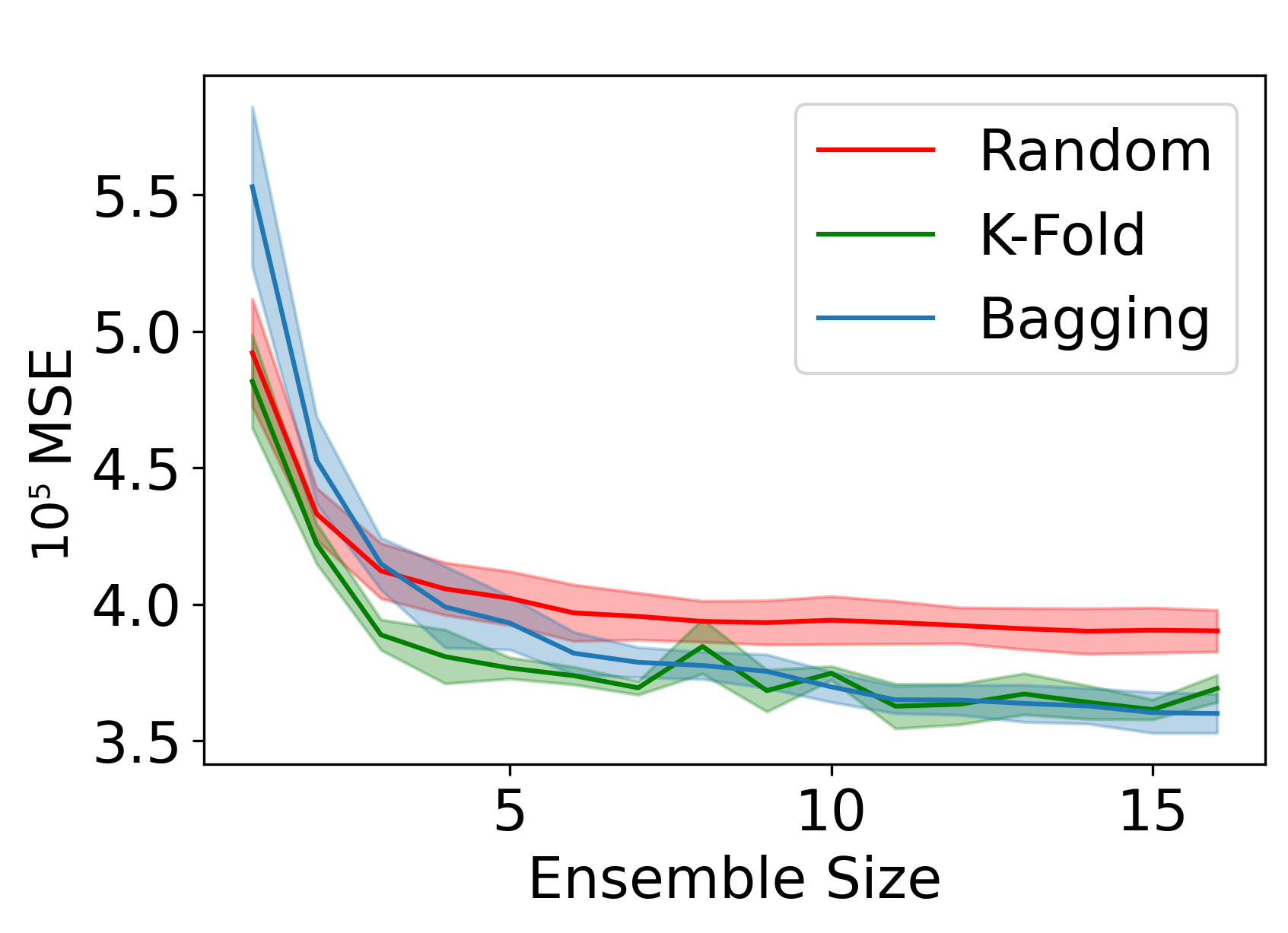}
 \caption{MSE vs. ensemble size for three ensemble methods: Random initializations (red), K-fold cross validation (green), bagging (blue) with standard deviation from five runs.
 }
\label{fig:mse_ens}
\end{figure}

Before exploring the quantified uncertainty using the 1D current profile data set, we first found the ensemble method which yielded the best MSE on the mean prediction. \autoref{fig:mse_ens} shows the MSE vs. ensemble size for the three ensemble methods. Bagging has a better MSE than random initialization and performs slightly worse than the K-Fold ensemble for small ensemble size.
Since bagging with 16 ensemble components yielded the smallest MSE, we evaluated this model's performance using the correlation plot and retention curve described in \autoref{sec:metrics}.

\begin{figure}[!h]
    \centering
    \includegraphics[scale=0.35]{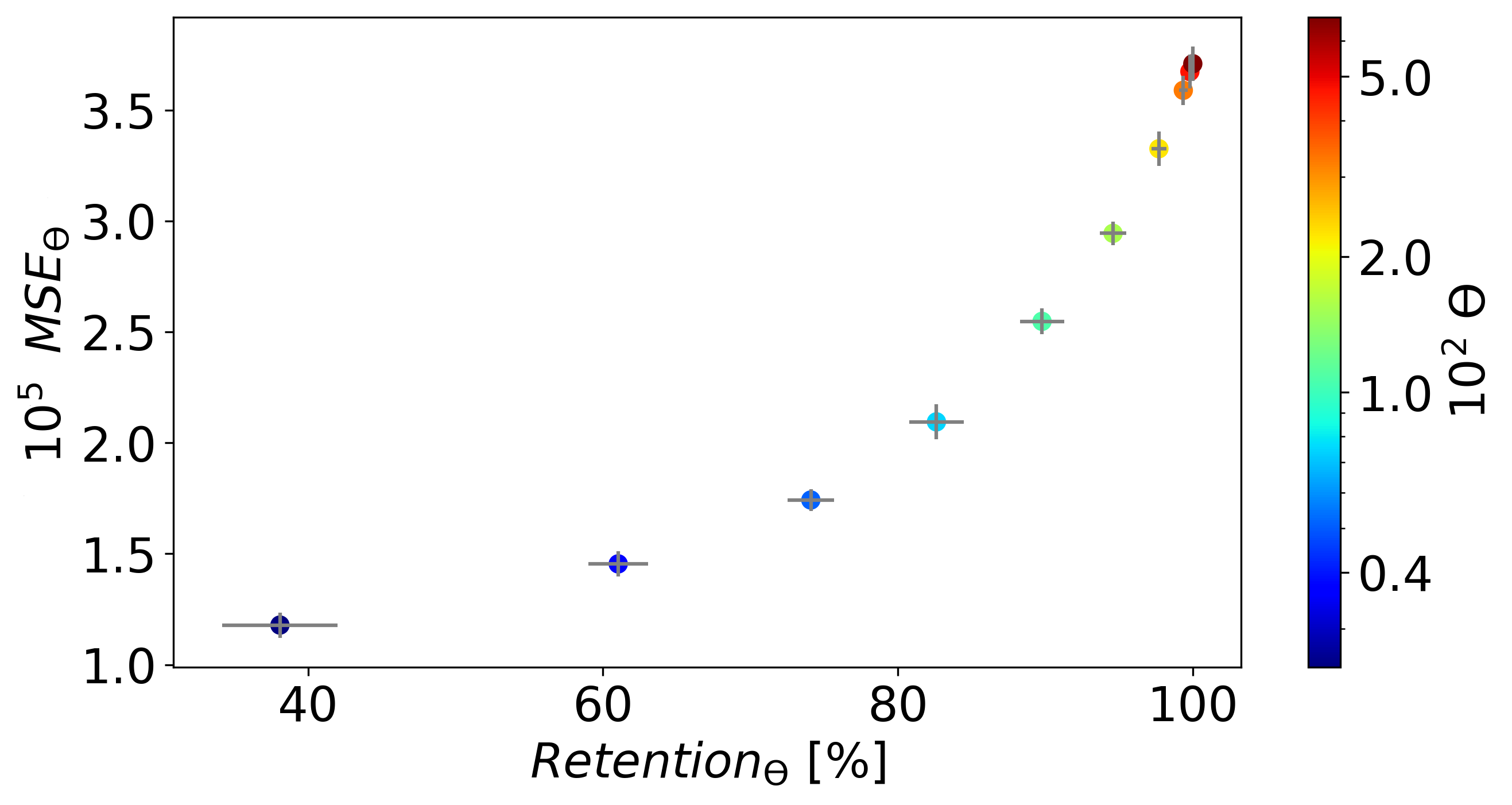}
    \caption{Retention Curve with the threshold ($\theta$) set by $\sigma_{\rm max}$ to evaluate the Bagging ensemble method. When a Retention$_{\theta}[\%]$ of the predictions are accepted, MSE$_{\theta}$ indicates the VD performance on that specific prediction subset.}
    \label{fig:retention}
\end{figure}

\begin{figure*}[!ht]
    \centering
       \begin{subfigure}[b]{.31\textwidth}
    \centering  \includegraphics[scale=0.3]{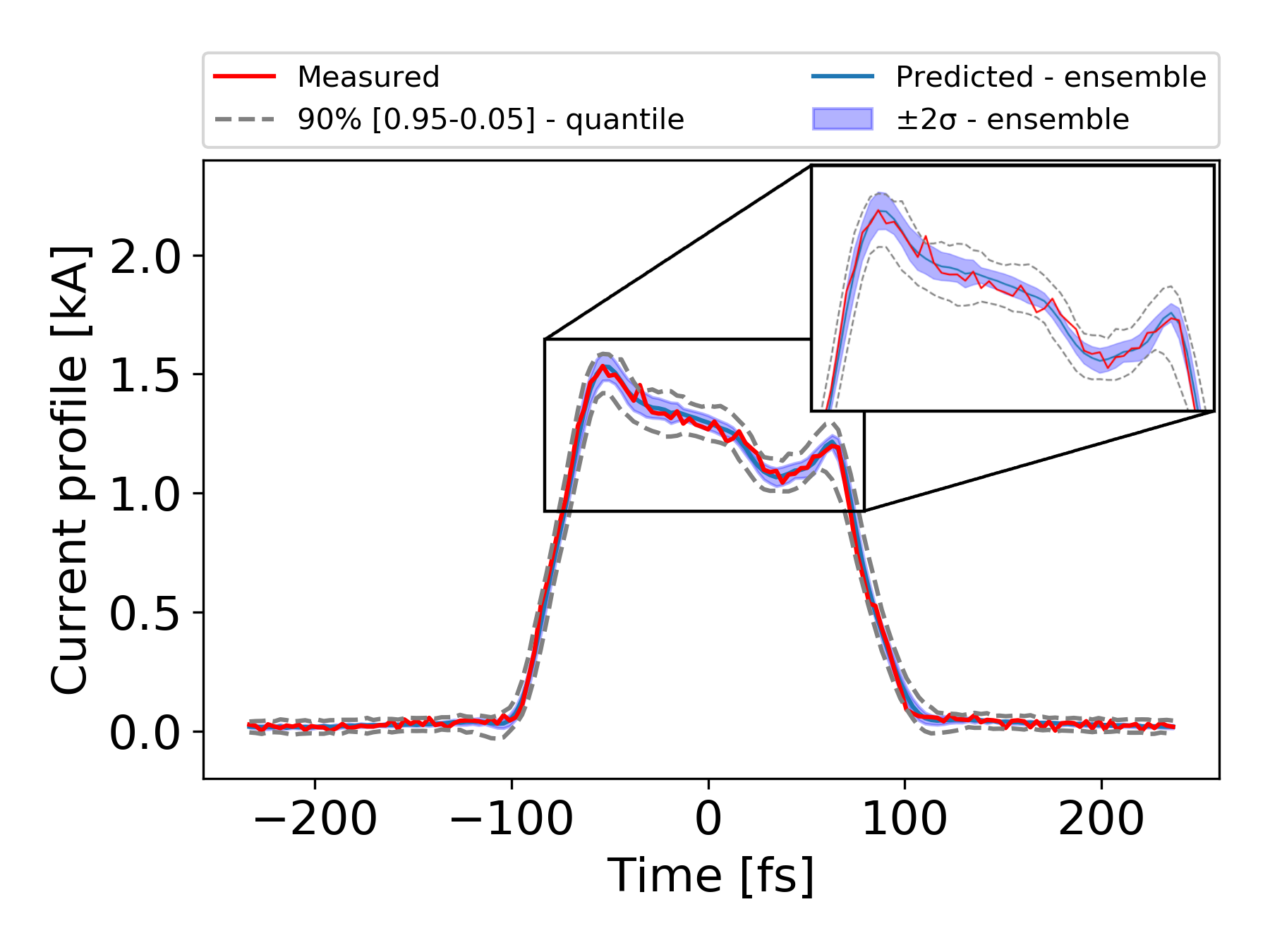} 
\caption{Test Shot} 
\label{fig:combo_id}
\end{subfigure}
\begin{subfigure}[b]{.33\textwidth}
    \centering  \includegraphics[scale=0.31]{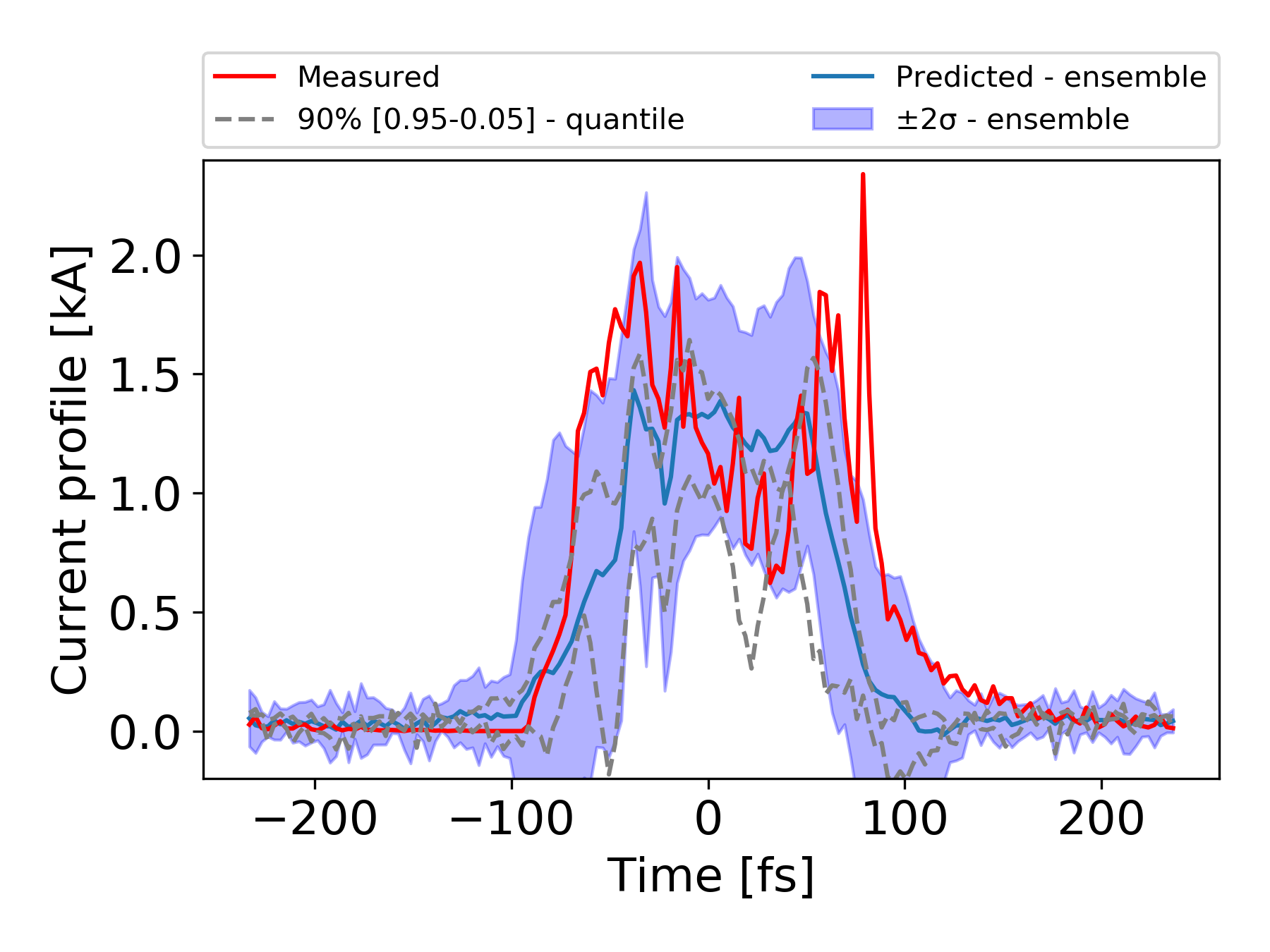} 
\caption{Out-of-distribution} 
\label{fig:combo_ood}
\end{subfigure}
    \begin{subfigure}[b]{.31\textwidth}
    \centering  \includegraphics[scale=0.3]{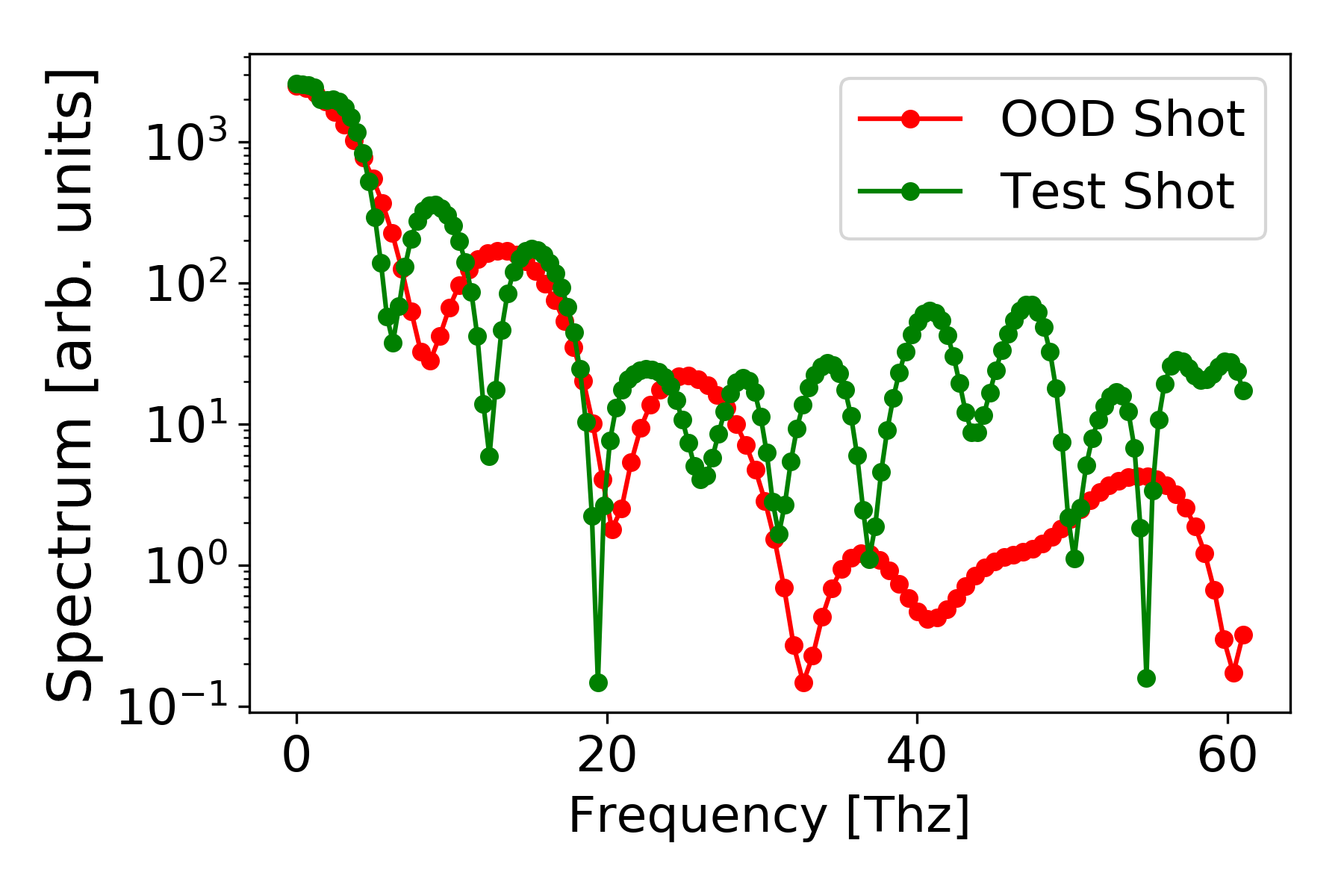} 
\caption{Spectra} 
\label{fig:combo_spec}
\end{subfigure}

    \caption{Comparison of (a) test shot in the same distribution as the training set, and (b) out-of-distribution (OOD) test shot. The uncertainty intervals are predicted with an ensemble model of Bagging with 16 components and 90\% quantiles (dashed). The spectra of both shots are compared in (c).}
    \label{fig:combo}
\end{figure*}

\autoref{fig:retention} shows retention curve, which can be used to set a rejection threshold for deployment of the VD.  Here, ${\rm MSE}_{\theta}$  is the average MSE for all test shots with $\sigma_{\rm max}$ lower than a given threshold $\theta$ displayed on the colorbar. 
The ${\rm Retention}_{\theta}$ on the x-axis describes the percentage of test shots we retain vs. reject in the process of choosing a threshold $\theta$. 
If a VD shot prediction yields a $\sigma_{\rm max}$ higher than the set $\theta$ value, we ignore the prediction. The retention curve is helpful for balancing the rejection/threshold trade-off for a specific use case and is comparable to a Receiver Operating Characteristic (ROC) curve used for binary classification tasks \cite{gneiting2018receiver}. In our case, we see a desirable positive correlation between ${\rm MSE}_{\theta}$ and ${\rm Retention}_{\theta}$. In deployment, the retention curve can be interpreted as follows: if the experimental conditions can  tolerate shots with averaged $\rm MSE_{\theta}\le2$e-5, by choosing $\theta=1.97$e-2, we can retain $80\%$ of the shots and yield a low $\rm MSE_{\theta}$ of 1.85e-05. 

Over the entire test set, the average MSE of bagging with 16 ensemble components was 3.65e-05 with an accuracy of 0.452 using $\pm1\sigma$ and 0.706 using $\pm2\sigma$ for computing $\alpha$ in \autoref{eq:accuracy}. The average MSE for quantile regression method (using $\tau=0.50$) was 4.05e-05. Accuracy was 0.973 using quantiles 0.05-0.95 and 0.648 using quantiles 0.25-0.75. \autoref{fig:combo_id} shows a test shot and the predicted uncertainty from the bagging ensemble ($\pm2\sigma$) and the $90\%$ of the quantiles to represent epistemic and aleatoric uncertainty respectively \cite{uq_comb_t}.

In order to evaluate the robustness of the model we use it to predict an out-of-distribution (OOD) sample. 
This means that the test shot has 
a different probability distribution than the training data set \cite{ovadia2019}. Therefore, it is expected that the model would manifest higher prediction uncertainty.
In particle accelerators, shifts in the distribution could come, for example, from a different operation mode or slow drift over time. In those cases, it is critical to have reliable predictive uncertainty. 
As an example, we compare a prediction of a test shot with the same distribution as the training data set (\autoref{fig:combo_id}), and a prediction for an out-of-distribution input from a different operation mode of the accelerator (\autoref{fig:combo_ood}). The spectrum inputs to the VD are shown in \autoref{fig:combo_spec}.
It is evident that the $\pm2\sigma$ interval from the ensembling method captures most of the uncertainty in the measured OOD shot. 

This shows that the VD can identify out-of-distribution shots, wherein the quantiles and ensemble predict abnormally large uncertainty ranges, and these shots can be rejected.
Previous works suggest that we can treat the predictive standard deviation from ensembles as epistemic uncertainty \cite{lakshminarayanan2017simple}, and quantile regression as doing maximum likelihood with an asymmetric Laplace distribution, hence capturing aleatoric uncertainty \cite{qr:laplace}. Further development of this technique would allow users to know whether the predicted uncertainty comes from lack of useful data, or variation within that training data.

\subsection{2D LPS Images\label{sec:lps}}

We used similar architecture to train ensembles with 2D longitudinal phase space images. The average MSE of the entire data set is 6.714e-04, and the accuracy with $\pm 1 \sigma$ is 0.291, and with $\pm 2 \sigma$ is 0.538. The accuracy of the image predictions was calculated using \autoref{eq:accuracy} evaluated across an additional dimension:
\begin{equation}
    {{\rm Accuracy }=  \frac{\sum_{t,e=1}^{T,E}\alpha_{t,e}\cdot  L_{{\rm measured},t,e}^{2}}{ \sum_{t,e=1}^{T,E} L^{2}_{{\rm measured},t,e}}}
        \label{eq:accuracy2D}
\end{equation}
where ${\rm \alpha}_{t,e} = \mathbbm{1}(| L_{{\rm measured}{,t,e}}| < L_{{\rm predicted}{,t,e}}\pm2\sigma_{t,e})s$.

\begin{figure*}[htpb!]
    \centering
    \begin{subfigure}{0.9\textwidth}
    \centering
    \includegraphics[scale=0.49]{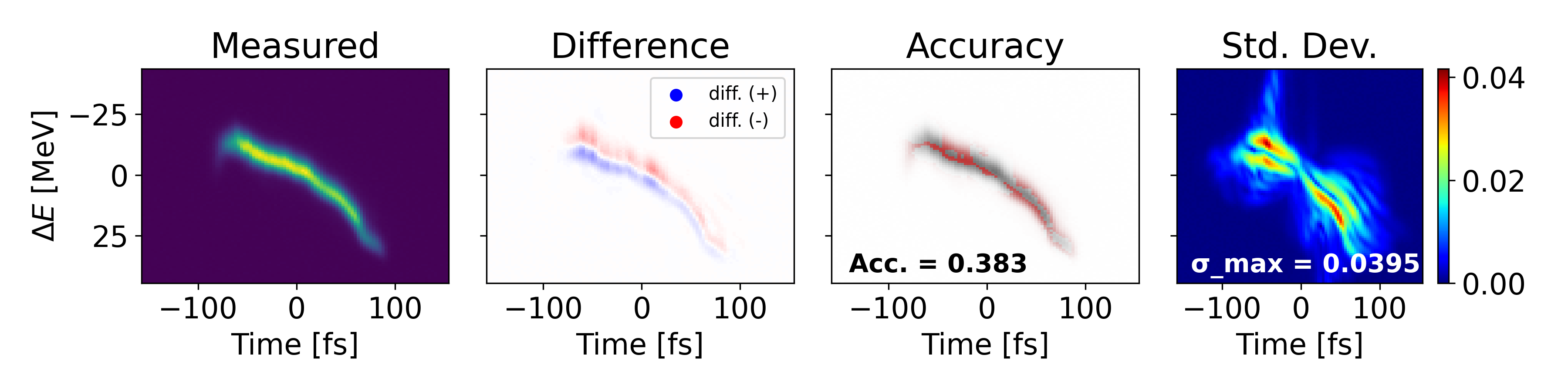}
    \caption{Translational error - Shot $\#762$}
    \label{fig:translational}
     \end{subfigure}
\begin{subfigure}{0.9\textwidth}
    \centering
    \includegraphics[scale=0.49]{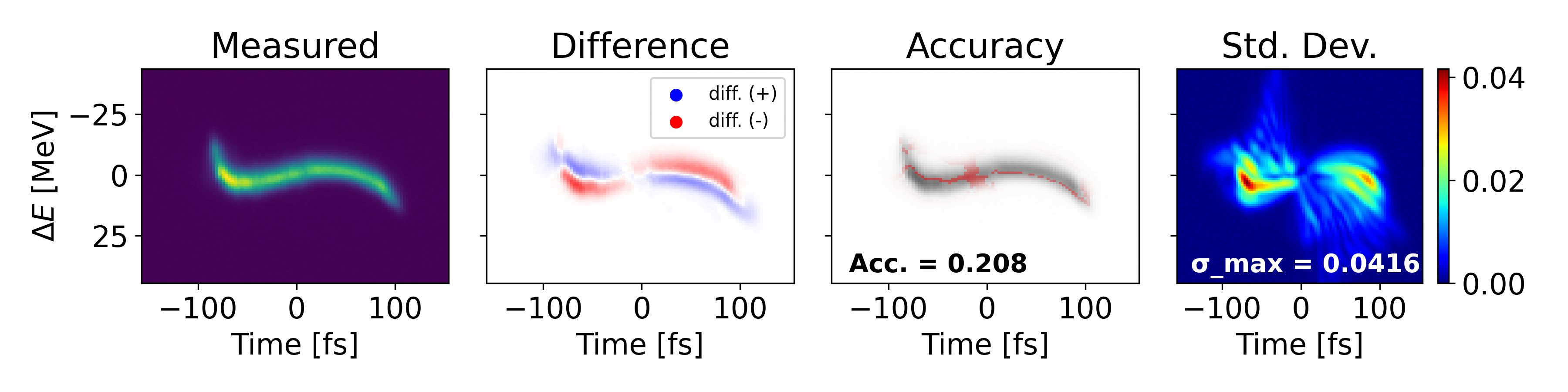}
      \caption{Shape error - Shot $\#789$}
    \label{fig:shape}
    \end{subfigure}
    \caption{2D longitudinal phase space (LPS) measurements with (a) translational error and (b) shape error examples.}
    \label{fig:imag}
\end{figure*}

\autoref{fig:imag} shows two examples of shots in the test set. The panels from left to right show the measured LPS image, the difference between the measured and predicted, the accuracy metric (Eq. \ref{eq:accuracy}), and the prediction uncertainty ($\sigma$). The blue and red colors on the difference panel indicates the positive and negative differences of the predicted LPS respectively. The red mask on the accuracy panel indicates regions where the measured value falls within the predicted uncertainty.

The two examples show predictions with translational error (shot $\#$762 in \ref{fig:translational}) and shape error (shot $\#$789 in \ref{fig:shape}) with corresponding MSEs of 2.737e-4 and 1.096e-3. We classified a prediction as one with translational error when the positive and negative differences are shifted with respect to each other.
For the translation error in shot $\#$762, shifting the mean prediction up by a single pixel brings a 85.0$\%$ decrease in MSE and a 41.8$\%$ increase in accuracy.
In order to determine how to shift the prediction, the `center of mass' of the prediction and the ground truth are calculated and then translated to match. In \autoref{sec:discuss}, we further discuss ways to leverage the spatial connectivity in these images to reduce such errors.


\section{Discussion\label{sec:discuss}}

In this section we discuss two techniques aiming to improve the LPS image neural network predictions. First we present the concept of a transposed convolution which can leverage the spatial connectivity in each data sample to produce better predictions. Then we present a specific ``Bottleneck" NN architecture that was applied and the insights derived from it. \newline

{\textbf{Transposed convolutions.}}
Transposed convolutions are a mathematical operation that allow for the up-scaling of data. Filters are trained to learn features which are then projected on to a feature map larger than the input. Transposed convolutions can be especially useful in regression tasks where the output involves important features and connectivity \cite{dumoulin2018guide}. The 2D outputs of the LPS image dataset are essentially images rather than vectors of unrelated measurements. Therefore, it makes sense to model architectures better fit for image generation which take advantage of the structured nature of the output. If we train filters to learn specific features in the data set (tails, curves, bright spots, etc.), the NN model would be better equipped to handle the data set at hand.\newline

{\textbf{Bottleneck architecture.}}  Bottleneck-shaped architecture is a NN design in which the number of neurons in the middle layer is smaller than in the other layers. The number of neurons per layer gradually decrease from the input to the middle layer, and then gradually increase from the middle layer to the output. The first part serves as an encoder or compressor which condenses essential information, and the second part serves as a decoder.
The middle layer allows for the input to be recorded as an $N$ dimensional vector. This embedding vector can be viewed as a representation of the instance in those $N$ dimensions, with similar data points having a small distance between them, and distinct ones being very distant.
Bottleneck architectures can reduce over-fitting by decreasing the system complexity \cite{Perin2020LearningWT}.

\begin{figure}[ht!]
    \centering
      \includegraphics[scale=0.5]{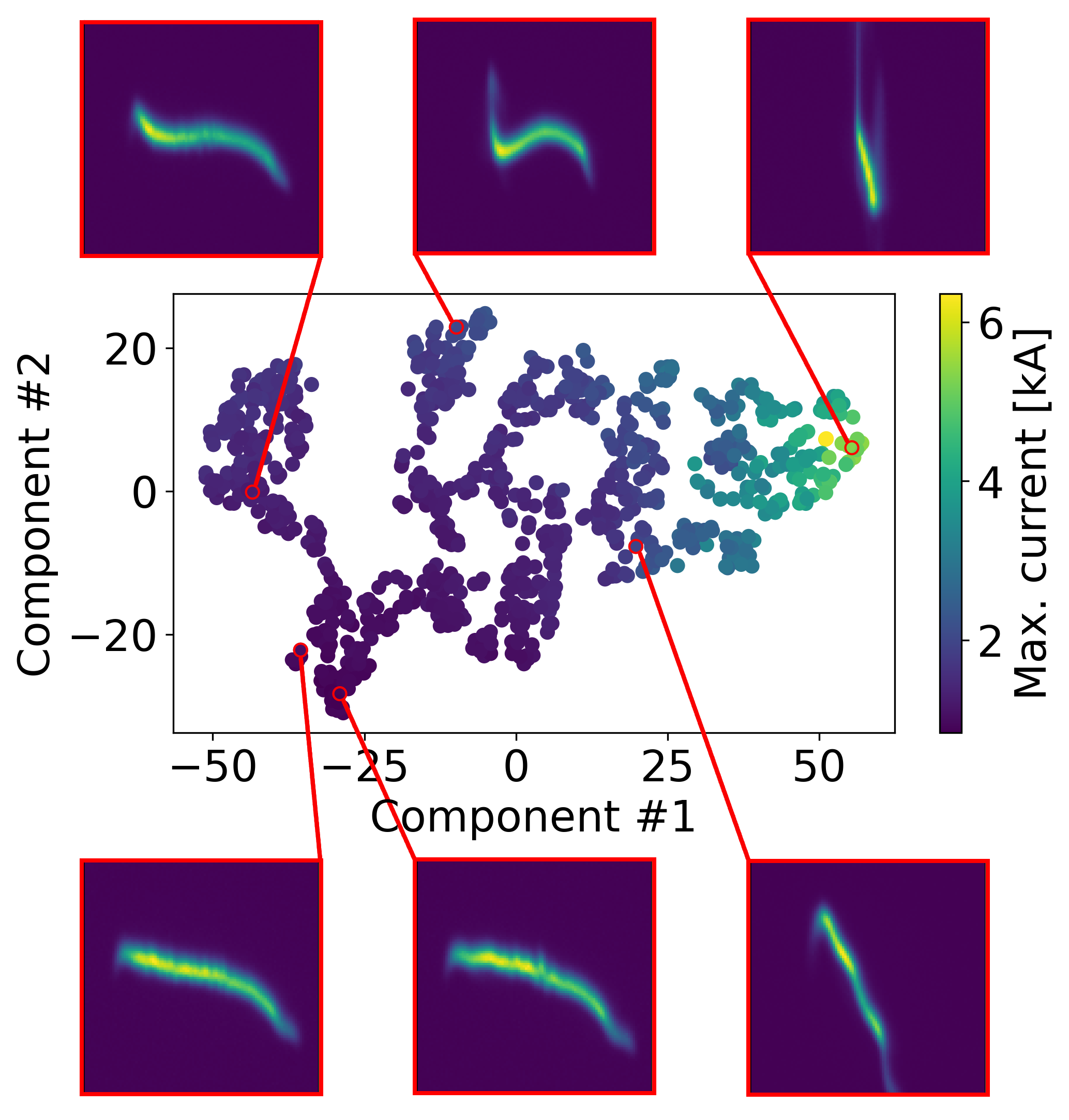}
 \caption{t-SNE clustering of the compressed vectors of the bottleneck architecture for all test shots, with six LPS examples of the outputs shown on the top. Notably, shots are grouped by shape and maximum current in the latent space.}
\label{fig:tsne}
\end{figure}

The encoder was a dense NN with layers of size 256, 128, and an embedding dimension of $N=64$. The decoder had transposed convolution layers of 8 2x2 filters, 8 4x4 filters, and 16 16x16 filters. Using this architecture, we trained a NN which achieved a negligibly improved average MSE of 8.833e-4. 
Nevertheless, a strength of the bottleneck architecture is its ability to condense information about the input and output relation into a single vector. In a high dimensional space, we expect that similar shots would aggregate close to each other, and distinctly different shots would be distant from one another. 

In order to visualize this, we used the t-SNE clustering algorithm \cite{TSNE} on all the embedding vectors for shots in the test set. After investigating multiple parameter values, we used a perplexity of $30$, a learning rate of $200$, and observed that the algorithm converged after $1000$ iterations. \autoref{fig:tsne} shows a visualization of the latent space of these vectors using two components. The maximum current for each shot is shown in color bar. The six examples shown illustrate that shots are grouped by shape and maximum current in the latent space.

\section{Conclusions \label{sec:conclusion}}

In this paper we presented several deep learning approaches to incorporate uncertainty in the virtual diagnostic (VD) tool. Deep learning models as neural networks have great predictive power, but they often suffer from over-fitting and provide over-confident predictions. 
Here, we compared three ensemble methods and quantile regression as a way to provide accurate mean predictions as well as correctly capturing confidence intervals.  Specifically, we considered a VD trained on non-invasive spectral measurements of the electron beam to predict the 2D longitudinal phase space or a 1D current profile.

The UQ methods presented in the paper were shown to be robust against out-of-distribution inputs by providing un-confident predictions on data it was not trained with. In comparison to a simple dense neural network, it was shown that a more tailored architecture can be used to exploit information about the data and offer better explainability during inference.
A principled approach to quantifying uncertainty is crucial for the deployment of the virtual diagnostic tool, especially for safety-critical systems as particle accelerators. \newline


\vspace{0.5cm}

\section*{Acknowledgments}

This work was supported by the Department of Energy, Laboratory Directed Research and Development program at SLAC National Accelerator Laboratory, under contract DE-AC02-76SF00515.

\appendix

\section{Re-Weighting the Current Profile Data Set\label{sec:reweighting}}

Upon further analysis of the data set, it became clear that certain current profile features are under-represented. As a result, the prediction of profiles with higher peak current presents high MSE. 
Typically, using models trained on this unbalanced data would be reasonable assuming nominal usage of the machine. However, although we have fewer examples of high peak shots, we would like to train the model to give them equal importance as the lower peak shots. 
Therefore, we weighted the output (current profile) such that we gave instances with lower representation more weight during training, and instances with over-represented samples less weight.
Re-weighting the data not only greatly reduced the MSE for many shots with high peak shots, but it also slightly reduced the overall MSE in other cases. 

In order to determine the weight for each shot, we made a 100-bin histogram of $I_{\rm max}$ for all $S=\sum_{i=1}^{100}n_i$ shots in the training set. Denoting $n_{i}$ as the number of shots in the $i$-th bin, the weight of the shots in that bin is ${ w_i } = c \cdot (n_{i})^{-0.5}$ with a normalizing constant ${ c } = S\cdot(\sum_{i=1}^{100}n_i^{0.5})^{-1}$. We tried other constants for the weighting power, but found that 0.5 as weighting constant performed the best.
The overall MSE of the test set was 4.601e-5 for the unweighted NN, and 4.478e-5 for the re-weighted NN (both averaged over 5 runs). 


\bibliographystyle{apsrev4-1}
\bibliography{ref}
\end{document}